# How to design a MAMS-ROCI (aka DURATIONS) randomised trial: the REFINE-Lung case study

**Authors:** Matteo Quartagno*^, Ehsan Ghorani**^, Tim P Morris, …, Michael J Seckl**^^, Mahesh K.B. Parmar*^^

**Affiliations:** *=Institute for Clinical Trials and Methodology, University College London, 90 High Holborn, WC1V 6LJ
^= Joint First authors.
^^= Joint Senior authors.
** Department of Medical Oncology, Charing Cross Hospital Campus of Imperial College London, UK.

**Corresponding author:** Matteo Quartagno, MRC Clinical Trials Unit, University College London, 90 High Holborn, Second Floor. WC1V 6LJ, London, UK.
Email: m.quartagno@ucl.ac.uk

**Abstract:**

**Background.** The DURATIONS design has been recently proposed as a practical alternative to a standard two-arm non-inferiority design when the goal is to optimise some continuous aspect of treatment administration, e.g. duration or frequency, preserving efficacy but improving on secondary outcomes such as safety, costs or convenience. The main features of this design are that (i) it randomises patients to a moderate number of arms across the continuum and (ii) it uses a model to share information across arms. While papers published to date about the design have focused on analysis aspects, here we show how to design such a trial in practice. We use the REFINE-Lung trial as an example; this is a trial seeking the optimal frequency of immunotherapy treatment for non-small cell lung cancer patients. Because the aspect of treatment administration to optimise is frequency, rather than duration, we propose to rename the design as Multi-Arm Multi-Stage Response Over Continuous Intervention (MAMS-ROCI).

**Methods.** We show how simulations can be used to design such a trial. We propose to use the ADEMP framework to plan such simulations, clearly specifying aims, data generating mechanisms, estimands, methods and performance measures before coding and analysing the simulations. We discuss the possible choices to be made using the REFINE-Lung trial as an example.

**Results.** We describe all the choices made while designing the REFINE-Lung trial, and the results of the simulations performed. We justify our choice of total sample size based on these results.

**Conclusions.** MAMS-ROCI trials can be designed using simulation studies that have to be carefully planned and conducted. REFINE-Lung has been designed using such an approach and we have shown how researchers could similarly design their own MAMS-ROCI trial.

# 1. Introduction

[NOTE: This is a work-in-progress paper and has not been submitted yet for consideration to any journal. If you had any comments, please contact the first author of the paper.]

Most clinical trials are designed with the goal to compare two different treatments, one of which generally represents the standard of care; the treatment variable is then binary in nature. However, in other situations, the research treatment might have to be chosen at a certain value on a continuum: one goal might be to find the optimal value for some aspect of treatment administration, like duration, dose or frequency of administration defined as the time between doses (henceforth simply referred to as "frequency"). This is often done at the early phases of drug development and the literature on dose-finding methods is vast; every new compound has to undergo phase 1 and/or phase 2 studies in order to select a dose to bring forward to later phases.

There are several cases, though, where optimisation of treatment administration might be required at later phases. In particular, this is true for some treatments already known to be effective, for which the evidence in favor of currently recommended guidelines is scarce or inconsistent, but early-phase trials can be skipped because it is known that some dose is better than none. The definition of 'optimisation' varies between applications: in some cases, we might be interested in simply maximising efficacy, but often the interest lies on identifying the least intense administration routine (the shortest duration/lowest dose/longest administration frequency) that leads to an acceptable treatment efficacy. In these situations, the goal is to try to reduce overall treatment administration compared to the current standard of care. There are several possible reasons to do so: it may reduce severe side-effects (e.g. for cancer immunotherapy), limit the spread of antimicrobial resistance (e.g. for antibiotics) or improve adherence (e.g. for novel TB treatments).

Since it is important to preserve efficacy while reducing overall treatment administration, the standard randomised trial design to address this problem has been a two-arm non-inferiority design comparing the current standard of care with a single reduced schedule.

One substantial issue with such design is that if the active arm (e.g., shorter duration, or lower dose than standard of care) is chosen poorly, the trial will inevitably give a negative result even though an optimal administration routine exists. The DURATIONS randomised trial design has been recently proposed[1] as a practical alternative in the specific settings of reducing antibiotic treatment duration; it tackles this issue by (i) randomising patients to multiple points along the continuum, increasing the chance to include the "optimal" research arm(s), and (ii) modelling the duration-response curve, sharing information across durations instead of treating them as independent, thus improving efficiency.

Standard frequentist trials aim to answer a simple binary question (e.g. is A non-inferior to B?) and are designed in order to control type I and type II errors in answering such a question. In most trials, type I error control is achieved by the analysis method, while type II error is controlled through design, under certain assumptions. The main question answered by DURATIONS trials is instead continuous in nature, as we already know that at least one duration, the standard one, is effective compared to no treatment, and we are mainly interested in finding the optimum in a pre-specified range. The relevant performance measures (i.e. the equivalents of power and type I error rate) are therefore less clear, complicating the design of a DURATIONS trial.

In (Quartagno et al, 2018), we designed a hypothetical trial focusing on a novel performance measure, the proportion of trials estimating the duration-response curve within a pre-specified absolute error. We used simulations to show that a sample size of 500 patients divided between 7 duration arms was enough to capture the duration-response relationship well, at least in the scenarios considered, that were motivated by an antibiotic trial, irrespective of the shape of the curve.

Such a performance measure would be suitable if the outcome of the trial was the whole curve. However, in practice, researchers would use that curve to make informed decisions. For this reason, we subsequently compared the properties of different methods of drawing inference from the duration-response curve. We found that a strategy based on the estimation of the shortest duration non-inferior to control within a certain margin was preferable, using non-parametric bootstrap to handle uncertainty.

Suppose a researcher has decided to design a DURATIONS randomised trial, with the plan to use the estimated response curve and the above analysis method in order to draw inference. How should they determine the most appropriate sample size, number and spacing of arms and other key design features?

This paper aims to answer this question using the REFINE-Lung trial as a case study. REFINE-Lung was designed to discover the optimally reduced frequency of immunotherapy administration for patients with advanced non-small cell lung cancer that is not less effective than the standard of care frequency. While the DURATIONS design was initially proposed with the motivation of reducing treatment duration, it can be applied to other situations with a continuous treatment variable where the goal is to find the optimal value for that variable. For example, it could also be used to optimise treatment administration frequency, as in our REFINE-Lung case study discussed in detail below. To reflect the flexibility of the approach to optimise variables beyond treatment duration, we hereafter refer to the design as Multi-Arm Multi-Stage Response Over Continuous Intervention (MAMS-ROCI) design. The originally described DURATIONS design is therefore a specific type of MAMS-ROCI where the treatment aspect to optimise is duration.

The rest of the paper is organised as follows: first, we introduce our case study, the REFINE-Lung trial; second, we give guidelines for planning a simulation study to determine the sample size required to adequately power a MAMS-ROCI trial; next, we present the design of REFINE-Lung, with the aim of guiding trial statisticians and researchers on ways to design MAMS-ROCI randomised trials more broadly. We conclude with a discussion and a plan for future work.

## 2. Motivating trial example: REFINE-Lung

Treatments that block immune inhibitory receptors (checkpoint immunotherapies; CPI) act to reinvigorate the immune response to control cancer. These therapies have transformed outcomes for patients with a range of cancer types. For example, in patients with advanced non-small cell lung cancer (NSCLC), 5 year survival is approximately 30-40% with the immunotherapy agent pembrolizumab compared to ~5-10% with chemotherapy, the previous standard of care.

The recommended dose of pembrolizumab is either 200mg every 3 weeks or 400mg every 6 weeks, but multiple lines of evidence suggest these dosing frequencies may be unnecessarily high, resulting in overtreatment (reviewed in the linked publication). Optimising administration frequency within a fixed time horizon is attractive given the high cost of the drug, quality of life implications associated with more frequent hospital visits for drug administration and the potential to reduce side effects.

However, it is not clear whether and by how much administration frequency can be reduced without significantly impacting efficacy. Thus, a MAMS-ROCI design is an appealing approach. The optimal dose frequency can be determined using a conceptually similar approach to determining the optimal duration of therapy as previously described. Patients are randomised to several treatment frequencies across a range, and the frequency-response curve is fitted to establish the optimal frequency that does not compromise efficacy vs. standard of care 6-weekly therapy.

Several practical questions arise: what should the total sample size be? Which frequencies should we patients be randomised between? Should we adopt an outcome-adaptive design? In the next following two sections we (i) illustrate possible design strategies and (ii) show how these were used to design the REFINE-Lung trial.

## 3. Design strategies

Suppose we want to design a MAMS-ROCI trial to optimise a certain aspect A of treatment administration that is a continuous variable. For example, we may want to optimise dose, frequency and/or duration of treatment. The currently recommended treatment is $A_{max}$, and we wish to investigate whether this can be reduced (or extended), to potentially as far as $A_{min}$. We originally proposed "500 patients in 7 arms" as a pragmatic sample size choice for estimating the treatment-response curve. This was calculated based on the specific example of a trial to reduce antibiotic treatment duration. However, sample sizes inevitably depend on design parameters specific to the question under investigation. For example, all treatment-response curves considered in (Quartagno et al, 2018)[1] assumed a very high cure probability in the control arm, as expected with antibiotic therapy for many infections. A suitable sample size calculation should be used in different situations, e.g. when the control event risk was expected to be much lower or higher than in our hypothesised scenarios. How should a researcher approach the sample size calculation and, more generally, the design, of a specific randomised trial?

Sample size calculations are usually based on analytic calculations. However, in the case of the MAMS-ROCI design, the use of flexible regression methods to handle a variety of plausible frequency-response curves makes analytic calculations more complex. Our preferred approach is therefore to use simulations to investigate different sample sizes. Recently, the ADEMP (Aims, Data generating mechanisms, Estimands, Methods and Performance measures) framework has been proposed as an approach to plan simulation studies to evaluate new statistical methods or to compare existing methods. We regard the ADEMP framework as the most useful way to plan simulation studies more generally, and so use it here to illustrate how to design a MAMS-ROCI study.

### 3.1. Aims

The aim is to determine the optimal sample size for a MAMS-ROCI trial. It is important to clarify which aspect of treatment one wishes to optimise (e.g. duration) and the upper limits of sample size feasibility.

### 3.2. Data-Generating Mechanisms

Standard sample size calculations for binary outcomes assume an expected event risk $\pi_{e,0}$ in the control arm and estimate the sample size needed to detect a statistically significant treatment effect with a given probability under the assumption that the intervention arm has a desired event risk $\pi_{e,1}$. In the case of conventional two arm non-inferiority trials, it is often the case that $\pi_{e,1} = \pi_{e,0}$, as the alternative hypothesis under which we want to be powered is that both arms are equivalent. One option for choosing the data-generating mechanism for our simulations is therefore to follow a similar strategy and generate the simulated data to model a flat treatment-response relationship, i.e. under the assumption that the treatment aspect to be optimised does not alter risk within the range considered. This is an appealing strategy from a patient perspective, as it implies that the goal is to identify treatments that are at least as effective as $A_{max}$ by making sure the trial is powered to extend frequency (or reduce duration) only when this is not at the cost of effectiveness.

An alternative is to generate data assuming a non-flat treatment-response relationship, based on a consideration of expert opinion and any other data sources that may be available. Under this approach, it is important that all arms are expected to be at least non-inferior to control, such that patients are not exposed to potential preventable harm.

Since the aim of the MAMS-ROCI design is to be as robust as possible to different possible shapes of the treatment-response curve, the above approach can be extended by simulating data under a set of plausible treatment-response scenarios, reflecting our uncertainty about the true relationship. For example, one possible option is to derive J scenarios, with J equal to the number of arms, in each of which $\pi_{e,0} = \pi_{e,j}$ but $\pi_{e,0} > \pi_{e,j+1}$. The advantage of this strategy is to ensure the trial is powered whatever the optimal arm is.

Other important choices in terms of data generating mechanisms are the number and position of arms and of the total sample size. For the sample sizes, a reasonable approach is to consider a range of sample sizes that reflect reasonable targets given available resources, as identified in the aims. A practical approach can involve choosing only multiples of 50, or 10 within the selected range. More efficient methods could also be used, for example following recommendations in (Wilson et al, 2021).

Regarding the choice of arms, the maximum should ideally represent the standard of care, while the minimum should be chosen with clinical considerations, as the most extreme one for which equipoise can be assumed.

Following (Quartagno et al, 2018), the safest and most general choice for the intermediate arms is to space them (approximately) equidistantly. The number of arms depends on the model used for the treatment-response curve. If using a fractional polynomial strategy with two power terms, at least 5 arms are needed, and (Quartagno et al, 2018) showed that it is unlikely that much can be gained beyond 7 arms, which appears to be an optimal choice in terms of power. Finally, practical and clinical considerations come into play as well since more arms for instance increases the complexity of the study.

### *3.3. Estimands*

The estimand represents the precise question to be addressed with our trial. The recent ICH E9 addendum[3] has given a new, more precise, definition of estimand which is made up of five attributes: (i) the treatment(s) to be compared, (ii) the population to be targeted, (iii) the primary outcome variable, (iv) the population-level summary measure (e.g. absolute risk difference vs risk ratio vs odds ratio) and (v) the way intercurrent events are handled.

In the original formulation of the MAMS-ROCI design, the overall goal was simply to estimate how the whole treatment-response curve would look like, while in (Quartagno et al, 2020)[2] we proposed alternative ways of drawing inference from the curve, i.e. different objectives that a MAMS-ROCI trial might try to achieve, where the aim was mainly to find the minimum non-inferior duration compared to control. The general features of these two possible estimands are listed in Table 1.

Note that the second estimand is broadly the one that would be targeted by a standard MAMS trial as well and the main differences between ROCI and standard MAMS trials are therefore in the analysis methods used to estimate such estimands, namely in the use of the model to share information across arms and in the strategy used to select the optimal duration, e.g. if picking the shortest duration non-inferior to control as the optimum.

### *3.4. Methods*

The first method choice is the model for the treatment-response curve. Depending on the outcome type and on the estimand, different models can be used, including logistic regression (for binary data) and Cox regression (for time-to-event data). Power of any method will depend on the optimal use of available information and on the choice of estimand.

For modelling the precise functional form of the treatment-response curve, fractional polynomial and spline-based methods are appealing choices, because of their flexibility and smoothness, which leads to greater efficiency than arm-specific contrasts against standard of care. However, more restrictive model choices may be preferable and more efficient when supported by available information. For example, Gompertz or logistic curves may be preferable if there is a good biological justification.

Whenever using one of the estimands presented in (Quartagno et al, 2020)[2], another method choice is that of the decision rule for identifying the optimal duration from the fitted curve. This is for example whether we want to identify the shortest duration non-inferior to control within a certain margin. Further, a third choice is that of the analysis method to account for uncertainty in the estimation of the population-level summary measure of interest. Bootstrap methods are a valid approach when using fractional polynomial models but may be too computationally expensive to be used for the whole simulation study. A good compromise might be to perform an initial sample size estimation using the delta-method approximation, and later evaluate only the selected sample size using bootstrap, possibly incrementing the estimated sample size by a certain amount, e.g. by 5% or 10%, to reflect the larger uncertainty expected from taking into account model selection variability.

### *3.5. Performance measures*

While in the design of standard 2-arm trials performance measures of interest are almost always power to reject the null hypothesis under the alternative, and type I error rate, in a MAMS-ROCI trial there is not a single hypothesis test against which these measures could be constructed. Rather, we *know* that the standard of care is superior to no treatment, and we want to investigate which arms are equivalent to the control one. We previously proposed three possible performance measures for MAMS-ROCI trials: (i) acceptable power, (ii) optimal power and (iii) type 1 error rate. Acceptable power is the probability that a trial ends up recommending an acceptable treatment different from the control. Optimal power is the probability that the trial recommends exactly the optimal treatment, for example the shortest one that is non-inferior to control. Type I error rate is the probability that the trial ends up recommending an insufficiently long, or frequent, treatment.

Targeting the usual 80% or 90% levels for optimal power can be a good choice under a flat curve scenario but may be too ambitious under a different model. The choice between the two measures may hence be guided by feasibility considerations and by the aim of the trial. For example, in certain situations reducing treatment duration by some amount may be considered a success even if the final duration recommended was not necessarily the optimal one.

Optimal and acceptable power are two extreme definitions of power but by no mean the only possibilities: all intermediate definitions may be considered, such as the probability of shortening treatment by at least 50% or by a certain fixed amount.

### *3.6. Other considerations*

There are additional considerations in the design of a trial that could be made but that don't easily fit into any of the above headings. One important example is whether we want to include adaptive elements into the design. This can be especially important for MAMS-ROCI trials, where it may be considered unethical to open all arms from the beginning of the study. While it could be possible to tweak the data-generating mechanism, estimand and methods sections of the ADEMP to include description of possible adaptive steps, it may be better at times to describe this separately, particularly if the adaptive step involves an interim analysis that is possibly quite different from the final planned analysis.

# 4. Implementation: REFINE-Lung

In this section we present the design of REFINE-Lung, using the ADEMP simulation framework just proposed.

## *4.1. Aim*

The aim of REFINE-Lung is to optimise the frequency of pembrolizumab administration among patients with advanced NSCLC. The treatment aspect to optimise is therefore frequency, and the goal is to seek extended frequencies non-inferior to the currently recommended 6-weekly standard. Given the population of interest, the number of potential centres and the timeline for the trial, sample sizes between 1000 and 2000 would be feasible.

## *4.2. Data-generating mechanisms*

The standard frequency is 6-weekly, with 18-weekly considered the most extended frequency patients could be recruited to. This choice was made based on a consideration of the likely maximum period of pembrolizumab effectiveness from phase I data. As many patients receiving immunotherapy treatment receive concurrent chemotherapy given on a 3 weekly administration schedule, to maximise the advantages of immunotherapy frequency reduction from a patient perspective, it is reasonable to restrict arms to frequencies that are multiple of 3. Hence, we consider five arms: 6, 9, 12, 15 and 18-weekly.

In terms of the frequency-response curve, we want to power the trial to find frequencies equivalent to 6-weekly; therefore we simulate the data assuming all frequencies are equivalent, i.e. under a flat frequency-response curve.

Additionally, we want to evaluate whether type I error rate is controlled, and hence we consider the 5 scenarios in Figure 1, where 18-weekly is always at the limit of non-inferiority.

### *4.3. Estimands*

Treatment is pembrolizumab administered every 6 to 18 weeks and the population id patients with NSCLC who were on treatment with no progression for six months. The primary outcome is 2-year overall survival; this was chosen as possibly the most clinically significant outcome, and was preferred to a time-to-event outcome is absence of a reasonable expectation that the hazards would be proportional. We choose risk ratio as population-level summary measure, i.e. we seek the most extended frequency non-inferior to 6-weekly within a certain margin defined on the risk ratio scale. Aggregating the results of the pembrolizumab trials in NSCLC patients published recently[4–6], a risk ratio margin of 0.88 would guarantee 50% preservation of effect.

The main intercurrent event to be expected is patients re-escalating to the standard dose on progression. As our approach is to evaluate the effectiveness of a strategy of administration of pembrolizumab, rather than the efficacy of the various frequencies, our main analysis will use a treatment policy strategy for handling this event, and so the analysis will include all randomised individuals.

### *4.4. Methods*

We use modified fractional polynomials of power 2 as the analysis method, where we select the best fitting among the 36 possible FP2 functions. The optimal frequency is dfined as the most extended one that is non-inferior to control within a 0.88 risk ratio margin. We initially estimate the optimal frequency using the delta-method approximation for computing confidence intervals for all the possible sample sizes. Having chosen a suitable sample size, we then check this choice by repeating the simulation where tests are constructed using bootstrap. We choose a one-sided significance level of 5%.

### *4.5. Performance measures*

We want to achieve 80% optimal power under the flat frequency-response curve. In other words, in the event all frequencies are equivalent, we want to have 80% probability of declaring 18-weekly as the optimal frequency. Additionally, we aim to control type I error rate at 5% under the alternative scenarios.

### *4.6. Other considerations*

As the currently recommended frequency is 6-weekly, and as pembrolizumab is an extremely effective drug for patients with NSCLC, immediately opening all 5 frequency arms may be seen as unethical because, although we assumed a flat frequency-response curve, the true relationship is unknown. We therefore introduce an adaptive element into the design. Initially we open the 6 and 12-weekly arms only, and we later proceed to open all the other arms if there is no significant reduction in efficacy with 12 weekly therapy in an interim analysis. Since 2-year overall survival would take too long to measure for an interim analysis, we base this test on progression-free survival and use a Cox model.

We plan to do this interim analysis within approximately 18 months from start of recruitment, in order to guarantee that only a certain fraction of total recruitment target in these arms would be achieved by that time. We compare the power of such a comparison being made after a certain number of events in the control arm have been achieved, under different assumptions for the experimental arm event risk. Note that, when implementing such analysis, a binary indicator of trial phase would have to be included to account for the different randomisation arms before and after the interim analysis.

### *4.7. Final design*

Figure 2 shows the results of the simulations run according to the above specifications. From this figure, a sample size of approximately 1550 patients appears enough to reach approximately 80% optimal power. However, this is based on the delta-method approximation, which ignores model selection variability. Using bootstrap to validate the selected sample size, gives slightly lower estimated power (72.6%, Monte-Carlo 95%CI: [68.7%,76.5%]). Increasing the sample size by 10% appears to be adequate (estimated power 79% [75.4%,82.6%]), so that the final sample size is 1750 patients.

Among the four type I error scenarios, the error rate is always less than 5%.

Figures 3-5 shows power of different sample sizes for detecting loss of effectiveness of the lower frequency arm at the interim analysis, using various design parameters. A sample size of approximately 150 patients appears to be enough to reach 80% power to detect a difference of 20 percentage points between the two arms, using a 5% two-sided confidence interval for the hazard ratio for progression-free survival.

## 5. Discussion

We have presented a simulation-based strategy for designing trials that use the MAMS-ROCI design. We have applied the proposed strategy to illustrate the design of the REFINE-Lung trial, whose aim is to optimise immunotherapy treatment frequency for patients with NSCLC.

### *5.1. Recommendations*

1. When the goal of a researcher is to optimise some aspect of treatment administration that can be seen as continuous, i.e. duration, dose or frequency, a MAMS-ROCI trial can be considered as a valid option;

2. It is important to explain and justify all choices, in particular that of the estimand.

3. When using a fractional polynomial model with 2 powers, at least 5 arms should be used (ideally 7), and these should be equally spaced whenever practically possible;

4. Once these choices have been made, a good way to calculate the necessary sample size is to perform a simulation study using the ADEMP framework such as the one presented here;

## *5.2. Extensions*

While the current methodology has been developed for binary outcomes, it can be readily extended to continuous and time-to-event data. The model currently fitted does not assume monotonicity of the curve. As in certain settings it is reasonable to assume more extended frequencies will not lead to better survival, it would be interesting to investigate the impact of restricting the models to force monotonicity.

Another plan is to investigate how to incorporate additional variables in the design, either simply as adjustment variables to gain power or with interactions, to perform subgroup analyses and possibly give different recommendations to different groups.

Finally, Bayesian methods could be considered particularly appealing for MAMS-ROCI designs, as the focus is more on modelling than on answering a single binary question and they could allow to compute the posterior probability of any point being non-inferior,. Possible extensions to design a MAMS-ROCI trial in the Bayesian framework might follow.

## *5.3. Other approaches*

Designing trials with simulations has two potential limitations: the first is that sample size calculations are valid only if the true duration-response mechanism is among the ones considered in the data-generating mechanisms. However, this is also true of most analytic sample size formulae for standard two-arm designs.

The second limitation is that the simulations may require very long running times, particularly if, as here, the analysis method involves bootstrap re-sampling and estimation of fractional polynomial models on each re-sample.

Alternative solutions could be sought to contrast this issue. A formal sample size calculation formula is difficult to derive, when the analysis model chosen is fractional polynomials, because it should somehow reflect model selection uncertainty. However, as our focus is on predicting mean outcomes rather than in estimation of model parameters, an alternative solution could be to adopt methods for sample size calculations in prediction studies[7,8], perhaps adapting them to the slightly different aims of our design.

When designing a clinical trial, it is common to have a maximum practicable sample size $N_{MAX}$, to maintain time and economical resources within feasible limits. A possible design approach is to investigate power of our MAMS-ROCI trial with sample size $N_{MAX}$ under a range of plausible duration-response curves. As there is not a single null hypothesis to test, the idea of recruiting as many patients as possible is justifiable, as any additional patients will give us more information about our duration-response curve.

Hence, one could run the simulations on the same scenarios as in 2.2, but only for sample size $N_{MAX}$, and explore power of the trial under different models. If this was at acceptable levels, a trial with sample size $N_{MAX}$ would be justified.

### *5.4. Conclusions*

The MAMS-ROCI, aka DURATIONS, randomised trial design is a novel multi-arm alternative to standard two arm non-inferiority designs for trials that seek to optimise duration, dose or frequency of treatment administration. We have shown that such a trial can be designed using simulations and illustrated the design of the REFINE-Lung trial, that can act as an example for future trials adopting the same methodology.

**Funding**

This work was supported by the Medical Research Council [MC_UU_12023/29].

**ORCID iD**

Matteo Quartagno https://orcid.org/0000-0003-4446-0730

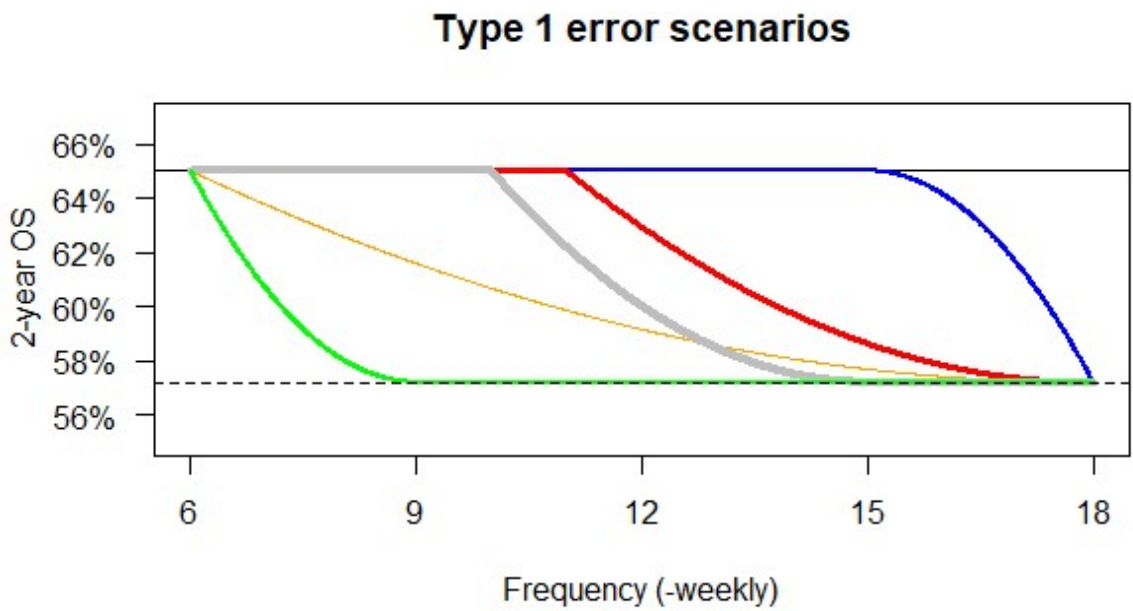

**Figure 1: Type I error scenarios investigated. 2-y OS at the control arm is always 65%, and in the 18-weekly arm it is always 88% of 65%, i.e. it lies on the non-inferiority margin.**

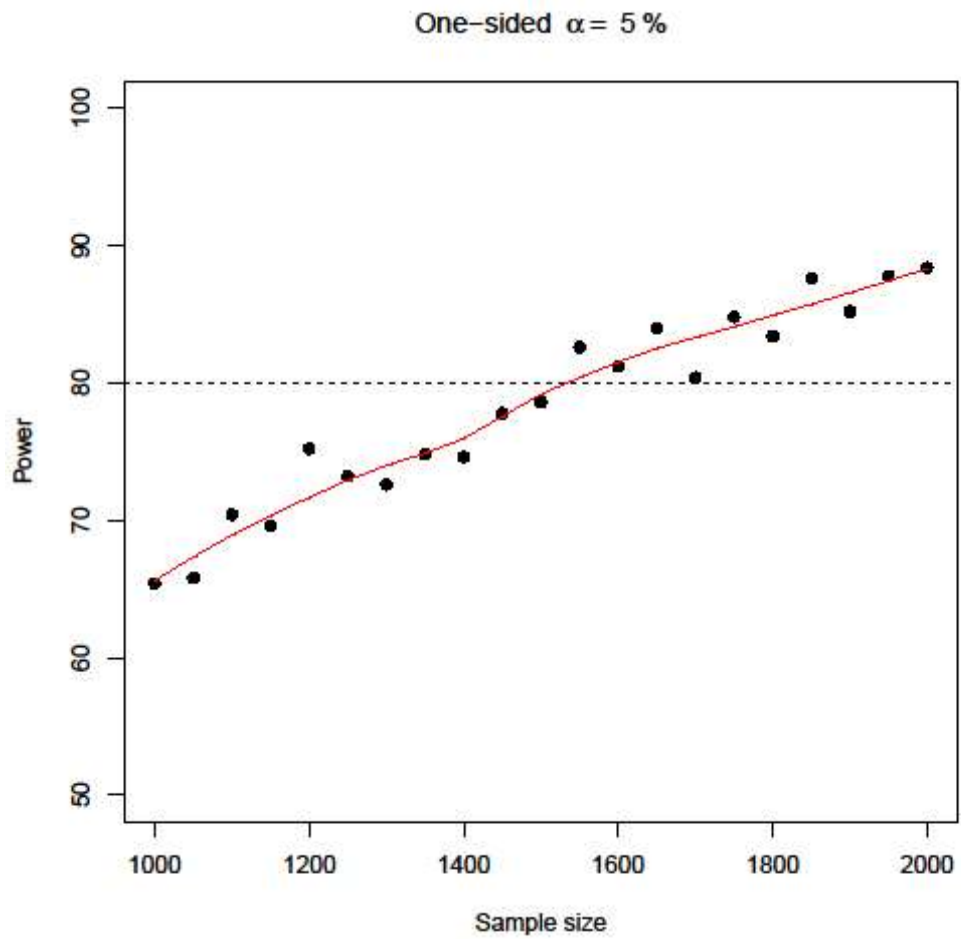

**Figure 2: Optimal power for different sample sizes between 1000 and 2000. The red lie is the loess smoother fitted through the estimated data points.**

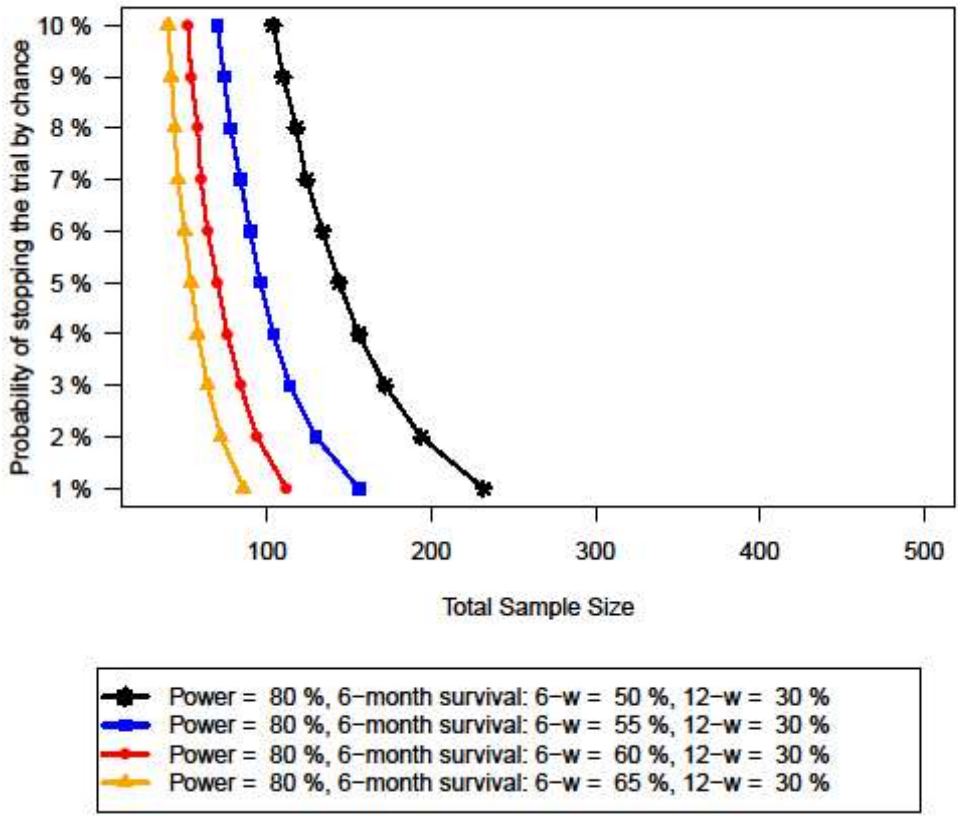

**Figure 3: sample sizes for the interim analysis varying significance level and 6-weekly (control) event rate, but keeping power fixed at 80% and experimental (12-weekly) event rate at 30%.**

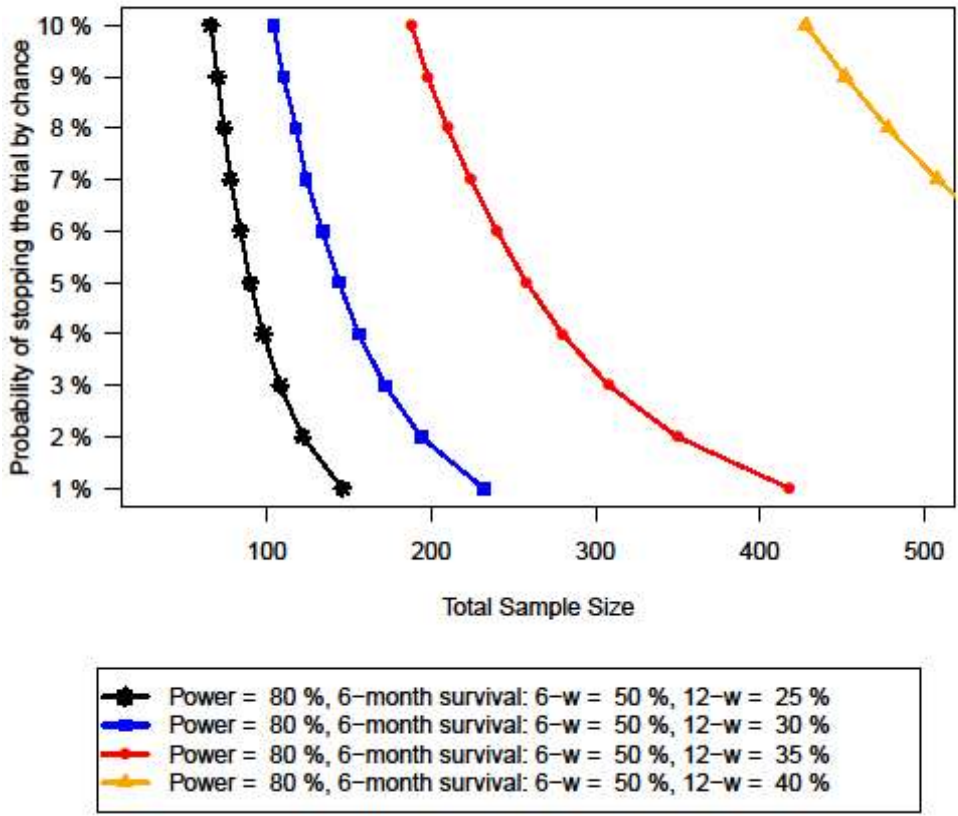

**Figure 4: sample sizes for the interim analysis varying significance level and 12-weekly (experimental) event rate, but keeping power fixed at 80% and control (6-weekly) event rate at 50%.**

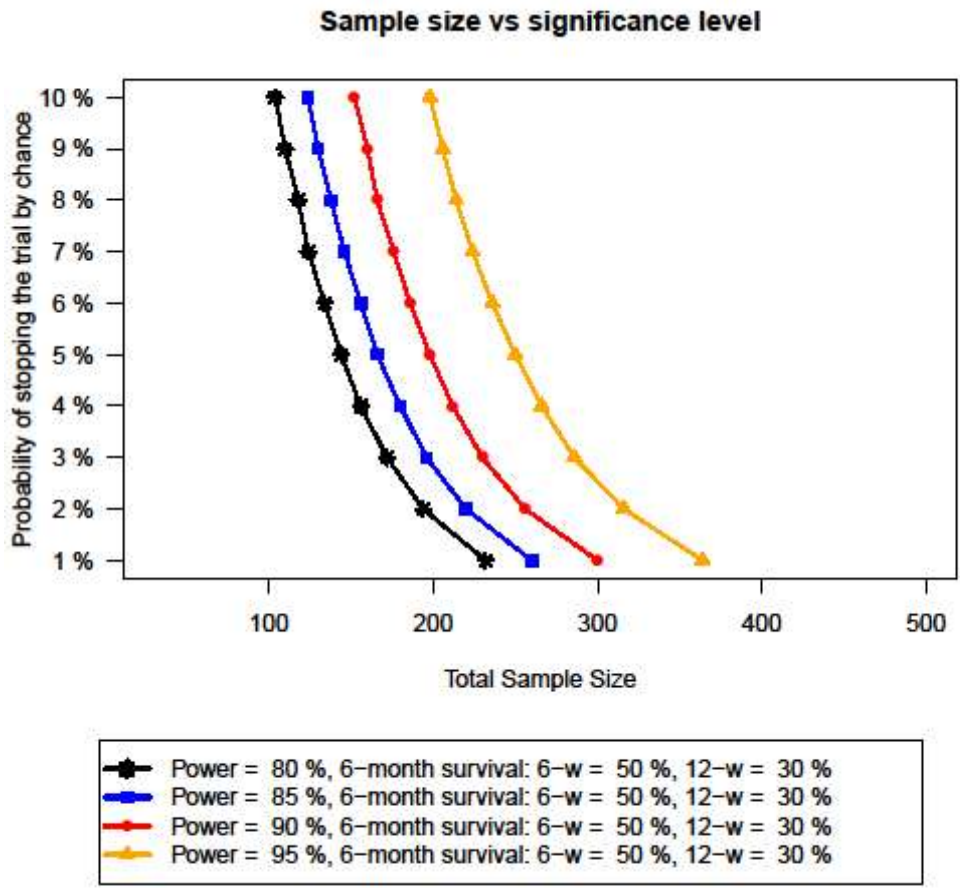

**Figure 5: sample sizes for the interim analysis varying significance level and power, but keeping control (6-weekly) event rate at 50% and experimental (12-weekly) event rate at 30%.**

**Table 1: Estimands used in the two methodological papers where the MAMS-ROCI design was proposed. Differences between the two estimands are highlighted in bold.**

| Attribute | 2018 paper estimand (whole treatment-response curve) | 2020 paper estimand (shortest duration non-inferior to control) |
|---|---|---|
| Treatment | Single standard of care treatment, at multiple doses/frequencies/durations, **but not compared against each other;** | Single standard of care treatment, at multiple doses/frequencies/durations, **comparing all against control dose/frequency/duration;** |
| Population | Target population of interest | Target population fo interest |
| Variable | Binary outcome variable | Binary outcome variable |
| Population-level summary measure | **Outcome risk at each point on the curve** | **Risk difference / ratio for each point vs longest duration** |
| Intercurrent events handling | Treatment policy strategy | Treatment policy strategy |